\documentclass[sigconf]{acmart}

\usepackage[utf8]{inputenc} 
\usepackage[T1]{fontenc}    
\usepackage{hyperref}       

\usepackage{hyperxmp}
\usepackage{url}            
\usepackage{booktabs}       
\usepackage{amsfonts}       
\usepackage{nicefrac}       
\usepackage{microtype}      
\usepackage{xcolor}         

\usepackage{multirow}
\usepackage{subfigure}

\usepackage[english]{babel}
\usepackage{moresize}
\usepackage{amsmath}
\usepackage{algorithmic}
\usepackage{balance}
\usepackage{comment}
\usepackage{paralist}
\usepackage{bm}
\usepackage{pgfplots}
\usetikzlibrary{pgfplots.dateplot}

\usepackage{flushend}
\usepackage[english]{babel}
\usepackage{graphicx}

\usepackage{amssymb}
\usepackage{amsfonts}
\usepackage{url}
\usepackage{bbm}
\usepackage{longtable}
\usepackage{rotating}
\usepackage{multirow}
\usepackage{mathrsfs}
\usepackage{enumitem}
\usepackage[linesnumbered,algoruled,boxed,lined]{algorithm2e}
\usepackage{adjustbox}
\usepackage{hyperref}
\usepackage{pgfplots}
\usetikzlibrary{pgfplots.dateplot}
\usepackage{filecontents}

\newcommand{\eg}{\textit{e}.\textit{g}.} 
\newcommand{\wrt}{\textit{w}.\textit{r}.\textit{t}}

\def\model{MixRec}

\AtBeginDocument{%
  }

\copyrightyear{2025}
\acmYear{2025}
\setcopyright{acmlicensed}\acmConference[WSDM '25]{Proceedings of the Eighteenth ACM International Conference on Web Search and Data Mining}{March 10--14, 2025}{Hannover, Germany}
\acmBooktitle{Proceedings of the Eighteenth ACM International Conference on Web Search and Data Mining (WSDM '25), March 10--14, 2025, Hannover, Germany}
\acmDOI{10.1145/3701551.3703591}
\acmISBN{979-8-4007-1329-3/25/03}

\begin{document}



\title{MixRec: Heterogeneous Graph Collaborative Filtering}

\author{Lianghao Xia}
\affiliation{%
  \institution{University of Hong Kong}
  \city{Hong Kong}
  \country{China}
  }
\email{aka\_xia@foxmail.com}

\author{Meiyan Xie}
\affiliation{%
  \institution{South China University of Technology}
  \city{Guangzhou}
  \country{China}
  }
\email{myxie23@foxmail.com}

\author{Yong Xu}
\affiliation{%
  \institution{South China University of Technology}
  \city{Guangzhou}
  \country{China}
  }
\email{yxu@scut.edu.cn  }

\author{Chao Huang}
\authornote{Chao Huang is the Corresponding Author.}
\affiliation{%
  \institution{University of Hong Kong}
  \city{Hong Kong}
  \country{China}
  }
\email{chaohuang75@gmail.com}


\renewcommand{\shortauthors}{Lianghao Xia, Meiyan Xie, Yong Xu, and Chao Huang}

\begin{abstract}
For modern recommender systems, the use of low-dimensional latent representations to embed users and items based on their observed interactions has become commonplace. However, many existing recommendation models are primarily designed for coarse-grained and homogeneous interactions, which limits their effectiveness in two critical dimensions. Firstly, these models fail to leverage the relational dependencies that exist across different types of user behaviors, such as page views, collects, comments, and purchases. Secondly, they struggle to capture the fine-grained latent factors that drive user interaction patterns. To address these limitations, we present a heterogeneous graph collaborative filtering model \model\ that excels at disentangling users' multi-behavior interaction patterns and uncovering the latent intent factors behind each behavior. Our model achieves this by incorporating intent disentanglement and multi-behavior modeling, facilitated by a parameterized heterogeneous hypergraph architecture. Furthermore, we introduce a novel contrastive learning paradigm that adaptively explores the advantages of self-supervised data augmentation, thereby enhancing the model's resilience against data sparsity and expressiveness with relation heterogeneity. To validate the efficacy of \model, we conducted extensive experiments on three public datasets. The results clearly demonstrate its superior performance, significantly outperforming various state-of-the-art baselines. Our model is open-sourced and available at: \textcolor{blue}{\url{https://github.com/HKUDS/\model}}.
\end{abstract}



\maketitle

\section{Introduction}
\label{sec:intro}

Modern recommender systems aim to understand users' interests by analyzing their interactions with items, such as clicks, reviews, and purchases. Collaborative filtering (CF) frameworks play a crucial role in generating item recommendations in these systems~\citep{chen2020revisiting}. The core of most existing CF approaches lies in their ability to map users and items into a shared latent space, effectively capturing users' preferences through various learning techniques~\citep{wu2021towards, yang2022hicf, fu2020magnn}. Notable examples of these techniques include multilayer perceptron~\citep{he2017neural}, auto-encoder~\citep{sedhain2015autorec}, and graph neural networks (GNNs)~\citep{wang2019neural, wang2020multi, fan2022graph}.

In practical Collaborative Filtering scenarios, user interaction data can be characterized by two crucial dimensions: \emph{Behavior Heterogeneity} and \emph{Intent Diversity}. Firstly, the interactions between users and items inherently exhibit heterogeneity due to the presence of multiple behavior types \cite{chen2020efficient,xu2023multi}. For instance, in E-commerce platforms, users engage in various item-level behaviors such as clicks, reviews, collections, and purchases. Similarly, in video streaming sites, behaviors like watching, liking, and sharing can vary among users. These diverse behaviors reflect the varied preferences of users at a granular level. Secondly, the formation of user-item interactions is influenced by various latent intent factors \cite{wang2020disentangled,li2021disentangled}. For example, users may have specific intentions, such as shopping for family parties or making purchases when products are on sale. However, the entanglement among these diverse user intents presents challenges when modeling heterogeneous user-item interaction patterns, an area that remains largely unexplored in current multi-behavior recommender systems (\eg, MBGCN~\cite{mbgcn2020}, KHGT~\cite{xia2020knowledge}, SMRec~\cite{gu2022self}).

To enhance the expressiveness of collaborative filtering models and disentangle the hidden factors influencing user-item interactions, considerable research has focused on learning disentangled representations from interaction data. Notable studies include MacridVAE \citep{ma2019learning}, which uses a variational auto-encoder framework; the graph neural network-based DGCF \citep{wang2020disentangled}; and the geometry-based GDCF \citep{zhang2022geometric}. While these methods have proven effective, they often overlook the benefits of incorporating behavior heterogeneity to capture diverse user-item relationships. Despite the potential advantages of addressing intent disentanglement alongside multi-behavior modeling, the area of heterogeneous disentangled collaborative filtering remains largely unexplored.

However, this task presents non-trivial challenges: (i) The design of intent disentanglement must be approached with meticulous care to ensure that the resulting factorized representations effectively capture the expressive information pertinent to different types of user-item interactions. (ii) The limited availability of target behavior data (e.g., purchases, likes) poses a significant obstacle for most supervised multi-behavior recommender systems (e.g., MBGCN \cite{mbgcn2020} and KHGT \cite{xia2020knowledge}), which struggle to facilitate customized knowledge transfer between auxiliary and target behaviors due to the scarcity of labeled instances. Inspired by the remarkable success of Self-Supervised Learning (SSL), data augmentation emerges as a pivotal enabler for achieving accurate user representation learning under conditions of label scarcity. In order to harness the benefits of SSL for heterogeneous collaborative filtering with intent disentanglement, it is crucial to carefully adapt the SSL tasks, ensuring their adaptability and effectiveness in empowering disentangled contrastive learning to handle the diverse user-item interactions.

Considering the challenges outlined above, we propose a novel heterogeneous graph collaborative filtering model, referred to as \model, which addresses behavior heterogeneity and disentangled intent factors through adaptive multi-behavior data augmentation. In \model, we begin by encoding users' relation-aware latent intents using a parameterized heterogeneous hypergraph, generating factorized representations that are specifically relevant to a particular type of user-item interaction. Each type of hyperedge is carefully tailored to aggregate diverse intent embeddings while considering the disentangled global collaborative relationships. To enhance the generalization capacity for encoding personalized multi-behavior dependencies from sparse interaction data, we introduce a relation-wise contrastive learning paradigm. This paradigm effectively models behavior heterogeneity at both the node and graph levels, incorporating adaptive self-supervision. Within our \model, we integrate multi-behavior discrimination objectives into the adaptive contrastive learning process. This integration ensures that the factorized representations with intent disentanglement accurately reflect the heterogeneous context.

In summary, our work makes the following key contributions: \vspace{-0.05in}

\begin{itemize}[leftmargin=*]

\item \textbf{Problem}. We explore recommender systems from the new perspective of learning factorized representations pertinent to diver user-item interactions. By elucidating the intent disentanglement with interaction heterogeneity encoding, this work allows recommenders to capture fine-grained diverse preference of users. \\\vspace{-0.12in}

\item \textbf{Methodology}. We propose a novel heterogeneous graph collaborative filtering paradigm, \model, that uncovers latent intent factors in heterogeneous user-item interactions. \model\ comprises two key components: (i) a learnable intent disentanglement module based on a parameterized heterogenous hypergraph neural network to encode diverse latent factors, and (ii) a relation-wise contrastive learning model with hierarchical structures for adaptive data augmentation at node- and graph-levels, facilitating the extraction of both local and global disentangled factors. \\\vspace{-0.12in}

\item \textbf{Theoretical Analysis}. We include a theoretical discussion that validates the efficacy of our model, particularly highlighting the strengths of adaptive self-supervised data augmentation. We also provide a comprehensive rationale analysis for our heterogeneous hypergraph contrastive learning paradigm. \\\vspace{-0.12in}

\item \textbf{Experiments}. Our \model\ method is evaluated on multiple public datasets to show the obvious performance improvement compared with a variety of baselines. We conduct further analysis to validate the model robustness, efficiency and interpretability.

\end{itemize}


\section{Preliminaries}
\label{sec:model}

In our heterogeneous collaborative filtering scenario, the relationships between users ($u_i$, where $i$ ranges from 1 to $I$) and items ($v_j$, where $j$ ranges from 1 to $J$) are manifested through diverse interactions, such as clicks, reviews, and purchases. Specifically, we partition $K$ types of heterogeneous user-item interactions into the target behavior (e.g., purchase in e-commerce or like in a video streaming platform) and other auxiliary behaviors, such as {click, collect} and {watch, review}. To represent heterogeneous interactions, we define a three-way tensor $\mathcal{X}\in\mathbb{R}^{I\times J\times K}$. Each element $x_{i,j,k}=1$ indicates that user $u_i$ has interacted with item $v_j$ with the $k$-th behavior type, while $x_{i,j,k}=0$ represents unobserved interactions.

In practical scenarios, users exhibit diverse intentions when interacting with items in recommender systems, influenced by factors including explicit preferences, implicit interests, temporal dynamics, social influences, and contextual information. Therefore, disentangled representation learning for heterogeneous collaborative filtering becomes essential to capture and model these varied user intentions effectively and comprehensively. This approach enables a better understanding of distinct user preferences in multi-behavior patterns, ultimately improving recommendation accuracy.\\\vspace{-0.12in}

\noindent \textbf{Task Formulation}. With the foregoing definitions, the heterogeneous disentangled collaborative filtering can be formally stated as follows: \textbf{Input}. The set of observed user-item heterogeneous interactions, denoted as $\mathcal{X}$. \textbf{Output}. The learning function $f(\cdot)$, which makes predictions for unobserved user-item interactions of the target interaction type (such as purchase or like). The function simultaneously explores the heterogeneous user-item interactions and captures the underlying disentangled user intentions within the context of relation heterogeneity.
\vspace{-0.05in}
\section{Methodology}
\label{sec:solution}

\subsection{Multiplex Graph Relation Learning}
In our heterogeneous collaborative filtering, we utilize the heterogeneous interaction data $\mathcal{X}$ to create a multiplex interaction graph $\mathcal{G}$ that captures diverse relationships between users and items. Inspired by successful GNN-based representation learning methods \citep{yu2020graph, fu2020magnn, tang2024higpt}, we employ a relation-aware message passing approach.
\begin{align}
    \label{eq:gcn}
    \textbf{z}_{i,k}^{(u/v)} = \sum\nolimits_{j\in\mathcal{N}_{i,k}} m_{i,j}\circ \textbf{e}_{j},~~~~
    \bar{\textbf{z}}_i^{(u/v)} = \sum\nolimits_{k=1}^K \textbf{z}_{i,k}^{(u/v)}
\end{align}
The set of neighboring nodes for user $u_i$ and item $v_j$ under the $k$-th behavior type are denoted as $\mathcal{N}_{i,k}$ and $\mathcal{N}_{j,k}$, respectively. To address overfitting, we apply dropout operators, represented by $m_{i,j}$ and $m_{j,i}$, with binary values during the information aggregation process \citep{wu2021self}. The $\circ$ symbol denotes the broadcasting multiplication operation. The general node embeddings for $u_i$ and $v_j$, represented by $\textbf{e}_i$ and $\textbf{e}_j$ respectively, are initialized through random sampling. The type-specific behavior embeddings for $u_i$ and $v_j$ with behavior type $k$ are denoted as $\textbf{z}_{i,k}^{(u)}$ and $\textbf{z}_{j,k}^{(v)}$, respectively. The aggregated multi-typed behavior representations, $\bar{\textbf{z}}_i^{(u)}$ and $\bar{\textbf{z}}_j^{(v)}$, are generated using the sum-pooling operator. To enhance model efficiency \citep{he2020lightgcn,chen2020revisiting}, \model\ employs a compact design for graph message passing.

\subsection{Relation-aware Intent Disentanglement}
To gain insights into the diverse intentions that underlie heterogeneous user preferences in recommendation systems, we propose incorporating the disentanglement of latent factors alongside behavior heterogeneity across different types of user-item interactions. Drawing inspiration from the effectiveness of hypergraph structures in modeling high-order connectivity \citep{feng2019hypergraph, yi2020hypergraph}, we construct a heterogeneous disentangled hypergraph that integrates the diversity of user intents with multi-behavior user-item interactions.

In our approach, for each type of user-item interaction, we generate multi-channel hyperedges of size $E$, which represent the number of latent intents for user interaction preference. These hyperedges connect different user/item nodes through parameterized node-hyperedge dependency modeling, enabling the capture of relation-aware global collaborative relationships. In essence, we generate learnable hypergraph adjacency matrices, denoted as $\mathcal{H}^{(u)}_k\in\mathbb{R}^{I\times E}$ and $\mathcal{H}_k^{(v)}\in\mathbb{R}^{J\times E}$, which connect nodes and hyperedges under the $k$-th type of user-item interactions. This construction allows us to effectively model and explore the intricate relationships between users, items, and latent intents in a context-aware manner.
\begin{align}
    \label{eq:hyperedge}
    \mathcal{H}^{(u)}_k = \textbf{Z}^{(u)}_k \cdot \textbf{W}^{(u)\top}_k,~~~~~\mathcal{H}^{(v)}_k = \textbf{Z}^{(v)}_k \cdot \textbf{W}^{(v)\top}_k
\end{align}
The behavior-aware user embeddings, denoted as $\textbf{E}^{(u)}_k \in \mathbb{R}^{I\times d}$, and item embeddings, denoted as $\textbf{E}^{(v)}_k \in \mathbb{R}^{J\times d}$, are composed using the aggregated representations $\bar{\textbf{z}}_i^{(u)}$ and $\bar{\textbf{z}}_j^{(v)}$. These embeddings capture the specific behaviors associated with the $k$-th interaction type. The parameters $\textbf{W}^{(u)}_k$ and $\textbf{W}^{(v)}_k \in \mathbb{R}^{E\times d}$ represent the embedding matrices for the hyperedges related to the $k$-th interaction type. Each embedding vector in these matrices encodes the latent features corresponding to a specific user intent.

\begin{figure*}[t]
    \centering
    \includegraphics[width=\textwidth]{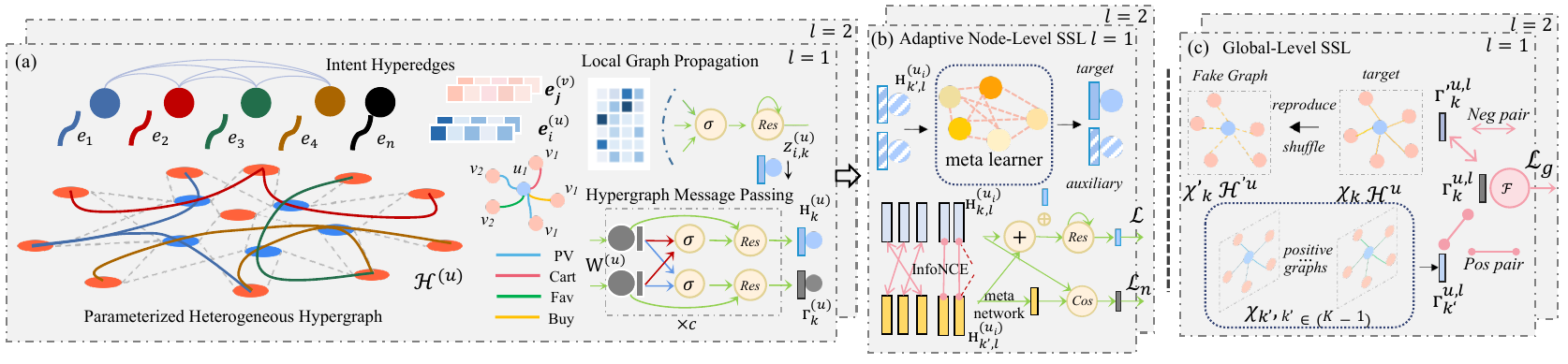}
    \vspace{-0.28in}
    \caption{\model's framework: (a) Our parameterized heterogeneous hypergraph enables relation-aware intent disentanglement, preserving behavior patterns. (b) Node-level contrastive learning captures local collaborative signals for cross-behavior knowledge transfer. (c) Graph-level contrastive learning performs data augmentation via multi-behavior user discrimination.}
    \vspace{-0.15in}
    \label{fig:framework}
\end{figure*}

\subsection{Parameterized Multi-Relational Hypergraph}
\subsubsection{\bf Hypergraph Structure Learning} With learnable hypergraph structures facilitating latent intent disentanglement, \model\ employs hypergraph-guided message passing to enhance user and item embeddings. This approach effectively captures user- and item-wise global collaborative relationships at fine-grained intent levels. The relation-aware hypergraph embedding propagation between leaf nodes and intent hyperedges is defined as follows:
\begin{align}
    \label{eq:hypergraph-guided}
    \textbf{H}_k^{(u/v)} = \delta(\tilde{\mathcal{H}}^{(u/v)}\mathbf{\Gamma}_k^{(u/v)}) = \delta(\tilde{\mathcal{H}}^{(u/v)} \delta(\tilde{\mathcal{H}}^{(u/v)\top} \textbf{Z}_k^{(u/v)}))
\end{align}
The matrices $\textbf{H}_k^{(u)}\in\mathbb{R}^{I\times d}, \textbf{H}_k^{(v)}\in\mathbb{R}^{J\times d}$ represent the smoothed node embeddings under the $k$-th behavior type for users and items, respectively. The non-linear activation function LeakyReLU is denoted by $\delta(\cdot)$. The embeddings of the $E$ hyperedges under the $k$-th behavior type for users and items are denoted as $\mathbf{\Gamma}_k^{(u)}$ and $\mathbf{\Gamma}_k^{(v)} \in \mathbb{R}^{E\times d}$, respectively. Here, $\mathbf{\Gamma}_k=\delta(\mathcal{H}^\top\textbf{Z}_k)$. In our disentangled hypergraph module, \model\ takes $\textbf{Z}_k^{(u)}\in\mathbb{R}^{I\times d}$ and $\textbf{Z}_k^{(v)}\in\mathbb{R}^{J\times d}$ (encoded from Equation~\ref{eq:gcn}) as input node embeddings. Through behavior-aware hypergraph message passing with intent disentanglement, \model\ captures the diverse intents behind multiplex user-item interactions, resulting in improved user representations. The overall user/item embeddings $\bar{\textbf{H}}$ are generated by sum-pooling the heterogeneous embeddings $\textbf{H}_k$.

\subsection{Hypergraph Embedding Propagation}
By combining local graph convolutions and global hypergraph message passing, \model\ applies the two modules alternately to recursively enhance user/item embeddings, injecting high-order information. Starting with the initialized node embeddings $\textbf{E}^{(u)}$ and $\textbf{E}^{(v)}$, denoted as 0-order embeddings $\mathbf{\Lambda}^{(u,0)}$ and $\mathbf{\Lambda}^{(v,0)}$, the $l$-th order node embeddings can be calculated iteratively as follows:
\begin{align}
    \mathbf{\Lambda}^{(u,l)} = \bar{\textbf{Z}}^{(u,l)} + \bar{\textbf{H}}^{(u,l)} +\mathbf{\Lambda}^{(u,l-1)} \nonumber\\
    \bar{\textbf{Z}}^{(u,l)} = f(\mathbf{\Lambda}^{(v,l-1)});~~~\bar{\textbf{H}}^{(u,l)}=g(\bar{\textbf{Z}}^{(u,l)})
\end{align}
The function $f(\cdot)$ represents relation-aware graph convolutions applied to the multiplex interaction graph $\mathcal{G}$. The function $g(\cdot)$ corresponds to the parameterized heterogeneous hypergraph neural network, which facilitates relation-aware intent disentanglement. To mitigate the problem of gradient vanishing, we incorporate a residual connection~\citep{he2016deep}. The final embeddings are obtained through a multi-order aggregation process, as outlined below:
\begin{align}
    \label{eq:aggregate}
    \mathbf{\Psi}^{(u)} = \sum\nolimits_{l=0}^L \mathbf{\Lambda}^{(u,l)},~~~~~ \mathbf{\Psi}^{(v)} = \sum\nolimits_{l=0}^L \mathbf{\Lambda}^{(v,l)}
\end{align}

\subsection{Multi-Relational Alignment with Adaptive Contrastive Augmentation}
Data sparsity is a common issue in heterogeneous collaborative filtering, where most users have limited supervision labels for behaviors like purchasing or liking. To enhance the model's capacity to capture diverse collaborative relationships, we developed a hierarchical contrastive learning paradigm utilizing hypergraph structures. This approach derives self-supervised signals from the original multi-behavior interaction data, incorporating both node-level and graph-level contrastive augmentation while preserving the dependencies among different behavior types. Our multi-relational hierarchical contrastive learning paradigm offers a comprehensive solution to the challenges of data sparsity and effectively encodes heterogeneous collaborative relations for recommendation.

\subsubsection{\bf Node-Level Adaptive Augmentation}
The diverse interactive patterns exhibited by users on online platforms result in varying dependencies among different types of behavior. For example, while some users may frequently favorite or like videos on TikTok, they may find it challenging to receive likes on the videos they have watched from others. To effectively capture these personalized multi-behavior patterns, our \model\ integrates a meta network encoder with self-supervised learning, enabling adaptive data augmentation. By leveraging multi-behavior self-discrimination, we model cross-type behavior relationships while preserving the personalized semantics associated with each behavior. To accomplish this, our meta network encoder is specifically designed to dynamically transform auxiliary behavior embeddings of different types, ensuring flexibility and adaptability in the encoding process.
\begin{align}
    \widetilde{\textbf{H}}^{(u,l)}_{i,k'} = {\textbf{H}^{(u,l)}_{i,k'}}\cdot{\delta{(\textit{Norm}(\textbf{H}^{(u,l)}_{i,k'}) \textbf{W}^{(u,l)}_{i,k'} + \textbf{b}^{(u,l)}_{i,k'})}}
\end{align}
The behavior-aware user embeddings, denoted as $\widetilde{\textbf{H}}^{(u,l)}_{i,k} \in \mathbb{R}^{I\times{d}}$, are adaptively transformed from the meta network for the $l$-th graph propagation layer. The activation function and the $l_2$ normalization function are represented by $\delta(\cdot)$ and $Norm(\cdot)$, respectively. The trainable parameters $\textbf{b}^{(u,l)}_{i,k'} \in \mathbb{R}^{d}$ and $\textbf{W}^{(u,l)}_{i,k'} \in \mathbb{R}^{d\times{d}}$ correspond to the $k'$-th type of auxiliary behaviors.

Our novel contrastive learning paradigm aims to provide additional supervision signals by strengthening user representations through multi-behavior self-discrimination. Specifically, based on the multi-view behavior embeddings of users, we consider the target behavior and each type of auxiliary behavior of the same user as positive pairs (i.e., ($\widetilde{\textbf{h}}_{i,k}, \widetilde{\textbf{h}}_{i,k'}$)). The embeddings of different users are treated as negative samples (i.e., ($\widetilde{\textbf{h}}_{i,k'}, \widetilde{\textbf{h}}_{i',k'}$)). Here, $k'$ represents the target behavior, and $k$ represents one auxiliary behavior. Within our contrastive learning paradigm, the augmented self-supervised signals enforce behavior divergence among different users. We employ InfoNCE to achieve node-level embedding agreement, as presented below:
\begin{align}
    \label{eq:node}
    \mathcal{L}_n = \sum_{i=1}^I\sum_{k=1}^K -\log \frac{\exp(s(\widetilde{\textbf{H}}_{i,k'}, \widetilde{\textbf{H}}_{i,k}) / \tau)}{\sum_{i'=1}^I\exp(s(\widetilde{\textbf{H}}_{i',k'}, \widetilde{\textbf{H}}_{i,k'} )/ \tau ) }
\end{align}
The node-level contrastive loss for the item dimension can be computed in a similar fashion. The function $s(\cdot)$ corresponds to the cosine function, which calculates similarity after normalizing the input vectors. The variable $\tau$ represents a temperature parameter that facilitates the learning of challenging samples during self-supervised learning. Through the augmented self-supervision provided by our node-level adaptive contrastive learning, our model gains the ability to effectively capture both the shared characteristics and the variations in multi-behavior preferences.

\subsubsection{\bf Graph-Level Adaptive Augmentation}
To incorporate global multi-relational context in our recommender system, we introduce graph-level contrastive learning with heterogeneous hypergraphs. Our parameterized component generates hyperedge embeddings by aggregating user information through learnable user-intent dependencies. This behavior-aware hyperedge representation serves as graph readout information, enabling intent disentanglement. In real-world scenarios, users with similar intents—such as prioritizing price or brand—may interact with items in various ways, including page views, purchases, or favorites.

With the aim of capturing the global multi-behavior context while considering the shared intent across behavior heterogeneity, we have designed a graph-level contrastive regularization for \model. To accomplish this, we first employ sum-pooling to aggregate the disentangled intent-aware information from all type-specific hyperedge embeddings. This aggregation process yields the behavior-specific hyperedge representation $\bar{\mathbf{\Gamma}}_{k}^{(u)}$ at the global graph-level. Subsequently, we consider the target-auxiliary behavior pairs as positive contrasting instances. Inspired by the negative sample generation strategy in graph infomax~\citep{velickovic2019deep,jing2021hdmi}, we construct graph-level negative pairs by randomly shuffling the learned hypergraph adjacent matrix, resulting in corrupted user embeddings $\bar{\mathbf{\Gamma}}_k^{'(u)}$. Formally, the graph-level multi-relational contrastive regularization is introduced through the following loss formulation:
\begin{align}
    \label{eq:graph}
    \mathcal{L}_g = -\sum\nolimits_{k} \log \frac{\exp(s(\bar{\mathbf{\Gamma}}_{k'},  \bar{\mathbf{\Gamma}}_k))}{\exp(s(\bar{\mathbf{\Gamma}}_{k'}, \bar{\mathbf{\Gamma}}_k)) + \exp(s(\bar{\mathbf{\Gamma}}'_{k'}, \bar{\mathbf{\Gamma}}_k))}
\end{align}
By employing the devised multi-relation contrastive learning, our \model\ approach utilizes adaptive contrastive self-supervised learning to enhance the modeling of multi-behavior dependencies derived from sparse and diverse interactions. This approach brings forth several advantages, including improved model robustness and enhanced recommendation performance.

\subsection{Model Optimization}
Utilizing the prediction score $\hat{\mathcal{X}}_{i,j,k} = \mathbf{\Psi}^{(u)\top}_i\cdot \mathbf{\Psi}^{(v)\top}_j$, we apply the pair-wise marginal loss function to the objective optimization. In particular, we randomly select $S$ positive and negative sample pairs for each user from their interacted and non-interacted items under the target behavior type. The overall optimized objective, which includes the recommendation loss, as well as the node-level and graph-level contrastive learning objectives, is defined as follows:
\begin{align}
    \label{eq:loss}
    \mathcal{L} & = \sum_{i=1}^N \sum_{s=1}^S \max(0, 1-(\hat{\mathcal{X}}_{i,p_s,k'} - \hat{\mathcal{X}}_{i,n_s,k'})) \\\nonumber & + \lambda_1 \cdot \|\mathbf{\Theta}\|_\text{F}^2 + \lambda_2 \cdot \mathcal{L}_n + \lambda_3 \cdot \mathcal{L}_g
\end{align}
The weights $\lambda_1, \lambda_2, \lambda_3$ are utilized to individually regulate the influences of weight-decay regularization, node-level contrastive regularization, and graph-level contrastive regularization. 
\\\vspace{-0.12in}

\noindent \textbf{Model Complexity Analysis}. We conducted a time complexity analysis of \model\ considering its three key modules: i) The complexity of multiplex graph relation learning is $O(|\mathcal{X}|\times d \times L)$, where $|\mathcal{X}|$ denotes the number of non-zero elements in $\mathcal{X}$. ii) The parameterized multi-behavior hypergraph neural network requires $O((I+J) \times E \times d)$ for message passing, where $E$ represents the number of latent factors for intent disentanglement. iii) The contrastive loss in the contrastive self-supervised learning component is calculated at the batch level, resulting in a complexity of $O(B^{(u)} \times B^{(v)} \times d)$, where $B^{(u)}$ and $B^{(v)}$ denote the number of users and items in a single batch, respectively. Based on this analysis, our \model\ exhibits competitive time complexity compared to some state-of-the-art GNN-based multi-behavior recommender systems.

\subsection{In-Depth Discussion of \model}
\subsubsection{\bf Adaptive Self-Supervision of \model}
We show that our heterogeneous hypergraph message passing mechanism not only generates additional supervision signals, but also provides learnable weights that facilitate adaptive CF training. Specifically, prediction scores $\hat{\mathcal{X}}_{i,j,k'}$ given by a vanilla GNN can be decomposed as:
\begin{align}
    \label{eq:gnn_decompose}
    \hat{\mathcal{X}}_{i,j,k'}^\text{G} = \textbf{z}_{i}^\top \textbf{z}_j
    =\sum\nolimits_{i'\in\mathcal{N}_i^L}\sum\nolimits_{j'\in\mathcal{N}_j^L} \alpha_{i'}\alpha_{j'}\cdot \textbf{e}_{i'}^\top\textbf{e}_{j'}
\end{align}
where $\alpha_{i'}, \alpha_{j'}$ represent coefficients related to node degrees in the heterogeneous interaction graph. Compared to vanilla GNNs shown above, our \model\ generates more supervision signals with learnable strengths coefficients, improving the parameter learning. Specifically, the prediction scores of \model\ can be decomposed:
\begin{align}
    &~~~~~~~\hat{\mathcal{X}}_{i,j,k'}^\text{H} = \textbf{H}_{i}^\top \textbf{H}_j = \left(\sum_{e=1}^E\delta\left(\tilde{\mathcal{H}}_{i,e}\cdot\sum_{i'=1}^I\delta\left(\tilde{\mathcal{H}}_{i',e} \cdot\textbf{e}_{i'}\right)\right)\right)^\top \nonumber\\
    &\cdot \left(\sum_{e=1}^E\delta\left(\tilde{\mathcal{H}}_{j,e}\cdot \sum_{j'=1}^J\delta\left(\tilde{\mathcal{H}}_{j',e} \cdot\textbf{e}_{j'}\right)\right)\right)
    =\sum_{i'=1}^I \sum_{j'=1}^J \beta_{i'} \beta_{j'} \cdot \textbf{e}_{i'}^\top \textbf{e}_{j'}
\end{align}
Assuming $\delta(\cdot)$ as the identity function, the high-order embeddings $\textbf{H}i$ and $\textbf{H}j$ can be associated with coefficients $\beta{i'}$ and $\beta{j'}$. In contrast to $\alpha_{i'}$ and $\alpha_{j'}$ used in vanilla GNNs, the weights $\beta_{i'}$ and $\beta_{j'}$ have two distinct characteristics: i) they are non-zero for all nodes in the graph, extending beyond the $L$-hop neighborhood, and ii) they are adaptively learnable, enabling the \model to optimize across heterogeneous graph structures.

\subsubsection{\bf Rationale of Graph Multi-Relational CL}
The graph-level contrastive learning objective of \model\ is able to adaptively and efficiently maximize the cross-relation similarity between nodes. This adaptive maximization is based on the strengths of nodes' connections to global hyperedges, which reflect the global connectivity of the nodes. Specifically, using dot-product as the similarity measurement $s(\cdot)$, we have the following simplified loss:
\begin{align}
    &\mathcal{L}_g = \sum_{k=1}^K -\bar{\mathbf{\Gamma}}_{k'}^\top \bar{\mathbf{\Gamma}}_k + \log\left(\exp(\bar{\mathbf{\Gamma}}_{k'}^\top \bar{\mathbf{\Gamma}}_k) + \exp(\bar{\mathbf{\Gamma}}_{k'}^{'\top} \bar{\mathbf{\Gamma}}_k)\right)
\end{align}
where $-\bar{\mathbf{\Gamma}}_{k'}^\top \bar{\mathbf{\Gamma}}_k$ and $\bar{\mathbf{\Gamma}}_{k'}^{'\top}\bar{\mathbf{\Gamma}}_k$ denote the positive and negative terms of contrastive learning, which can be further decomposed into:
\begin{align}
    -\bar{\mathbf{\Gamma}}_{k'}^\top \bar{\mathbf{\Gamma}}_k 
    &=-\sum_{e_1,e_2}\sum_{i_1,i_2}\tilde{\mathcal{H}}_{i_1,e_1,k'}\tilde{\mathcal{H}}_{i_2,e_2,k}\cdot \textbf{e}_{i_1}^\top \textbf{e}_{i_2}\nonumber\\
    \bar{\mathbf{\Gamma}}_{k'}^{'\top} \bar{\mathbf{\Gamma}}_k
    &=\sum_{e_1,e_2} \sum_{i_1,i_2} \epsilon \cdot \tilde{\mathcal{H}}_{i_2,e_2,k}\cdot \textbf{e}_{i_1}^\top \textbf{e}_{i_2}\nonumber
\end{align}
For positive samples, $\partial -\bar{\mathbf{\Gamma}}_{k'}^\top \bar{\mathbf{\Gamma}}_k / \partial \textbf{e}_{i_1}^\top \textbf{e}_{i_2} = \tilde{\mathcal{H}}_{i_1,e_1,k'}\tilde{\mathcal{H}}_{i_2,e_2,k}$ maximizes similarity between node pairs $(i_1, i_2)$ based on their relations to hyperedges. Additionally, $\partial -\bar{\mathbf{\Gamma}}_{k'}^\top \bar{\mathbf{\Gamma}}_k / \partial \tilde{\mathcal{H}}_{i_1,e_1,k'}\tilde{\mathcal{H}}_{i_2,e_2,k} = \textbf{e}_{i_1}^\top \textbf{e}_{i_2}$ adaptively maximizes cross-relation hypergraph structures for target relation $k'$ and auxiliary relation $k$ using node embedding similarity. For negative samples, one hypergraph weight is replaced with a noise coefficient $\epsilon$, resulting in similarity minimization with random strength. It achieves uniform contrastive optimization against similarity maximization for positive samples.

\vspace{-0.05in}
\section{Evaluation}
\label{sec:eval}

\begin{table}[t]
    \caption{Statistics of the experimented datasets}
    \vspace{-0.15in}
    \label{tab:data}
    \centering
    \footnotesize
	\setlength{\tabcolsep}{0.6mm}
    \begin{tabular}{ccccc}
        \toprule
        Dataset&User \#&Item \#&Interaction \#&Interactive Behavior Type\\
        \midrule
        Beibei&21,716&7,977& 3,338,068 &\{View, Cart, Purchase\}\\
        Tmall&114,503&66,706&5,751,432&\{View, Favorite, Cart, Purchase\}\\
        IJCAI&423,423&874,328&36,222,123&\{View, Favorite, Cart, Purchase\}\\
        \hline
    \end{tabular}
    \vspace{-0.15in}
\end{table}

In this section, we compare our \model\ to various baselines and aim to answer the following research questions:
\begin{itemize}[leftmargin=*]
\item \textbf{RQ1}: How does \model\ compare to state-of-the-art baselines? 
\item \textbf{RQ2}: What are the advantages of adaptive augmentation and intent disentanglement for heterogeneous collaborative filtering? 
\item \textbf{RQ3}: How robust is our \model\ model when confronted with sparse interaction data challenges within recommender systems?
\item \textbf{RQ4}: What is the impact of key parameters in \model\ model?
\item \textbf{RQ5}: Does \model\ offer interpretation analyses for global user dependencies and relation heterogeneity in recommendation?
\item \textbf{RQ6}: How does the efficiency of \model\ compare to baselines?
\end{itemize}

\subsection{Experimental Setup}
\noindent \textbf{Dataset}. We conduct experiments on three public datasets: Beibei, Tmall, and IJCAI. Table~\ref{tab:data} provides a summary of their characteristics. The \textbf{Beibei} dataset was collected from a popular e-commerce platform specializing in maternal and infant products. It contains 2,412,586 page views, 642,622 add-to-cart interactions, and 282,860 purchase interactions. The \textbf{Tmall} dataset serves as a benchmark for the recommendation task with heterogeneous interactions, encompassing four types of user-item interaction behaviors: page view (4,542,043), add-to-favorite (201,402), add-to-cart (516,117), and purchase (491,870). The \textbf{IJCAI} dataset, released by the IJCAI competition, which models user online activity from an online retailing site. It shares the same behavior types as the Tmall dataset: page view (30,287,317), add-to-favorite (2,934,022), add-to-cart (74,168), and purchase (2,926,616). In our heterogeneous collaborative filtering approach, we treat purchase behaviors as target interactions, while considering other types of interactions as auxiliary behaviors.

\begin{table*}[t]
    \centering
    \small
    \setlength{\tabcolsep}{0.6mm}
    \centering
    \caption{Overall performance comparison of all methods over different datasets in terms of \textit{HR@10} and \textit{NDCG@10}.}
    \label{tab:overall}
    \vspace{-0.12in}
    \begin{tabular}{c|c|c|c|c|c|c|c|c|c|c|c|c|c|c|c|c|c|c|c|c}
        \hline
        Data & Model & MF & NCF & CDAE & NADE & NGCF & SGCN & DGCF & GDCF & ~~ICL~ & NMTR & EHCF  & MBGCN & GNMR &  KHGT & MRec & MHCN & SGL & HCCF &
        \emph{\model} \\
        \hline
        \hline
        \multirow{2}{*}{Bei}
        & HR & 0.588 & 0.594 & 0.608 & 0.608 & 0.611 & 0.609 & 0.612 & 0.623 & 0.645 & 0.613 & 0.633 & 0.642 & 0.623 & 0.640
        & 0.610 & 0.614 & 0.619 & 0.610 & \textbf{0.679} \\
        \cline{2-21}
        & NDCG & 0.333 & 0.338 & 0.341 & 0.343 & 0.369 & 0.343 & 0.344 & 0.377 & 0.396 & 0.349 & 0.384 & 0.376 & 0.358 & 0.385 & 0.354
        & 0.345 & 0.346 & 0.354 & \textbf{0.419} \\
        \hline
        \multirow{2}{*}{Tma}
        & HR & 0.265 & 0.306 & 0.326 & 0.332 & 0.321 & 0.339 & 0.395 & 0.363 & 0.421 & 0.361 & 0.370 & 0.478 & 0.461 & 0.468 & 0.359
        & 0.411 & 0.413 & 0.374 &\textbf{0.525}\\
        \cline{2-21}
        & NDCG & 0.165 & 0.174 & 0.193 & 0.194 & 0.191 & 0.191 & 0.267 & 0.211 & 0.278 & 0.206 & 0.210 & 0.273 & 0.261 & 0.282 & 0.207
        & 0.249 & 0.261 & 0.215 & \textbf{0.325}\\
        \hline
        \multirow{2}{*}{IJC}
        & HR & 0.285 & 0.459 & 0.455 & 0.469 & 0.461 & 0.452 & 0.478 & 0.499 & 0.517 & 0.481 & 0.556 & 0.463 & 0.541 & 0.552 & 0.482
        & 0.504 & 0.484 & 0.487 & \textbf{0.611}\\
        \cline{2-21}
        & NDCG & 0.185 & 0.294 & 0.288 & 0.304 & 0.292 & 0.285 & 0.306 & 0.331 & 0.358 & 0.304 & 0.408 & 0.277 & 0.338 & 0.359 & 0.305
        & 0.332 & 0.316 & 0.317 & \textbf{0.418}\\
        \hline
    \end{tabular}
    \vspace{-0.08in}
\end{table*}

\subsubsection{\bf Baseline Methods} To comprehensively evaluate the effectiveness of our \model\ recommender, six groups of 18 baselines are included for recommendation performance comparison.

\noindent \textbf{1) Conventional Collaborative Filtering Models}
\begin{itemize}[leftmargin=*]
\item \textbf{MF}~\cite{koren2009matrix} and \textbf{NCF}~\cite{he2017neural}: These methods are the traditional and the deep neural versions of matrix factorization method, respectively.


\end{itemize}

\noindent \textbf{2) Autoencoder/Autoregressive Collaborative Filtering}
\begin{itemize}[leftmargin=*]

\item \textbf{CDAE}~\cite{wu2016collaborative} and \textbf{NADE}~\cite{zheng2016neural}: These two methods are bottleneck networks for user/item interaction vectors, with autoencoding and autoregressive training objectives, respectively.



\end{itemize}

\noindent \textbf{3) GNN-enhanced Recommender Systems}
\begin{itemize}[leftmargin=*]

\item \textbf{NGCF}~\cite{wang2019neural}: It utilizes recursive message passing in a GNN to capture high-order relationships among users and items.

\item \textbf{SGCN}~\cite{zhang2019star}: It employs GNNs with encoder-decoders to address sparsity by using embedding reconstruction loss with masking.

\end{itemize}

\noindent \textbf{4) Recommendation with Disentangled Representations}

\begin{itemize}[leftmargin=*]

\item \textbf{DGCF}~\cite{wang2020disentangled}: This approach encodes latent factors using GCN and partitions user representation into intent-aware embeddings.

\item \textbf{GDCF}~\cite{zhang2022geometric}: The model uses multi-typed geometries for interaction disentanglement and user factorized representations.

\item \textbf{ICL}~\cite{chen2022intent}: The method improves robustness by modeling latent intent distributions with contrastive self-supervised learning.

\end{itemize}

\noindent \textbf{5) Multi-Behavior Recommender Systems}

\begin{itemize}[leftmargin=*]

\item \textbf{NMTR}~\cite{gao2019neural}: This method captures correlations between different types of interactions between users and items using pre-defined cascaded behavior relations in a multi-task learning framework.

\item \textbf{EHCF}~\cite{chen2020efficient}: This approach correlates behavior-aware predictions for recommendation using a non-sampling transfer learning approach to tackle heterogeneous collaborative filtering.

\item \textbf{MBGCN}~\cite{mbgcn2020}: It models multi-typed user-item relations by propagating behavior-aware embeddings over a heterogeneous user-item interaction graph with a GCN backbone.

\item \textbf{GNMR}~\cite{xia2021multi}: It integrates self-attention and a memory network to encode behavior-specific semantics and dependencies, combining low-order user embeddings with high-order representations.

\item \textbf{KHGT}~\cite{xia2020knowledge}: This framework aggregates behavior-aware embeddings using a graph transformer network, distinguishing propagated messages with attentive weights.

\item \textbf{MRec}~\cite{gu2022self}: It models correlations between target and auxiliary behaviors using a star-style contrastive learning task.

\end{itemize}

\noindent \textbf{6) Self-Supervised Recommendation Models}

\begin{itemize}[leftmargin=*]
\item \textbf{SGL}~\cite{wu2021self}: It augments the user-item interaction graph with random walk-based node and edge dropout operators to construct contrastive views for self-supervised learning.

\item \textbf{MHCN}~\cite{yu2021self}: It incorporates a generative self-supervised task into the recommendation loss by maximizing mutual information between path-level and global-level embeddings.

\item \textbf{HCCF}~\cite{xia2022hypergraph}: It is a hypergraph CF model that generates self-supervised signals through local-global node self-discrimination.

\end{itemize}

\subsubsection{\bf Evaluation Protocols and Hyperparameter Settings}
To assess the accuracy of our recommendations, we use two common metrics: Hit Ratio ($HR@N$) and Normalized Discounted Cumulative Gain ($NDCG@N$). Following established practices~\cite{xia2020knowledge,mbgcn2020}, we employ a leave-one-out evaluation strategy, constructing the test set with users' last interactions under the target behavior type. In our \model\ model, we carefully select hyperparameters to ensure optimal performance. The number of latent intent hyperedges for each behavior type is chosen from {32, 64, 128, 256, 512}. During training, we explore different batch sizes (32, 64, 128, 256, 512) and dropout ratios (0.2, 0.4, 0.6, 0.8). The number of message passing layers in all graph-based methods is tuned from {1, 2, 3}. We also consider the regularization parameter $\lambda_{1}$ (range: $1e^{-5}$ to $1e^{-2}$) for $L2$ regularization. Furthermore, we fine-tune the regularization strengths $\lambda_2$ and $\lambda_3$ (range: $1e^{-7}$ to $1e^{-4}$) for the node-level and graph-level contrastive objectives, respectively. These comprehensive hyperparameter settings ensure the effectiveness of our model.

\begin{table}[t]
    \centering
    \small
    \caption{Performance with top-$N$ item recommendations.}
    \label{tab:topn}
    \vspace{-0.1in}
    \setlength{\tabcolsep}{1.9mm}
    \begin{tabular}{c|c|c|c|c|c|c}
        \hline
        Model & Metric & Top-1 & Top-3 & Top-5 & Top-7 & Top-9\\
        \hline
        \multirow{2}{*}{MF}
        & HR & 0.1184 & 0.3099 & 0.4526 & 0.5367 & 0.5795\\
        \cline{2-7}
        & NDCG & 0.1184 & 0.2275 & 0.2866 & 0.3164 & 0.3293\\
        \hline
        \multirow{2}{*}{NCF}
        & HR & 0.1228 & 0.3165 & 0.4471 & 0.5298 & 0.5796\\
        \cline{2-7}
        & NDCG & 0.1228 & 0.2316 & 0.2834 & 0.3154 & 0.3300 \\
        \hline
        \multirow{2}{*}{ICL}
        & HR & 0.1263 & 0.3170 & 0.4629 & 0.5437 & 0.5876\\
        \cline{2-7}
        & NDCG & 0.1636 & 0.3170 & 0.3552 & 0.3787 & 0.3887\\
        \hline
        \multirow{2}{*}{SGL}
        & HR & 0.1252 & 0.3218 & 0.4666 & 0.553 & 0.5949\\
        \cline{2-7}
        & NDCG & 0.1252 & 0.2357 & 0.2962 & 0.3288 & 0.3381\\
        \hline
        \multirow{2}{*}{EHCF}
        & HR & 0.1775 & 0.3977 & 0.5098 & 0.5759 & 0.6178 \\
        \cline{2-7}
        & NDCG & 0.1775 & 0.3029 & 0.349 & 0.3724 & 0.3887\\
        \hline
        \multirow{2}{*}{GNMR}
        & HR & 0.1395 & 0.3436 & 0.468 & 0.5458 & 0.6005\\
        \cline{2-7}
        & NDCG & 0.1395 & 0.2567 & 0.3086 & 0.3334 & 0.3515\\
        \hline
        \hline
        \multirow{2}{*}{\model}
        & HR & \textbf{0.1967} & \textbf{0.4153} & \textbf{0.5426} & \textbf{0.6132} & \textbf{0.6618}\\
        \cline{2-7}
        & NDCG & \textbf{0.1967} & \textbf{0.3224} & \textbf{0.3748} & \textbf{0.3992} & \textbf{0.4142}\\
        \hline
        \end{tabular}
        \vspace{-0.15in}
\end{table}

\vspace{-0.05in}
\subsection{Performance Comparison (RQ1)}
Based on the results presented in Table~\ref{tab:overall} and Table~\ref{tab:topn}, our \model\ approach demonstrates several key observations that highlight its effectiveness in capturing and leveraging heterogeneous behavior characteristics for accurate and robust recommendation.

\begin{itemize}[leftmargin=*]

\item Our proposed model surpasses existing baselines, underscored by an impressive maximum p-value of $1.6e^{-5}$. The exceptional performance of our approach underscores the merit of capturing diverse behavioral characteristics to construct a nuanced representation of user preferences. With the strategic implementation of a multi-behavior contrastive hypergraph module to deftly separate user intents, our model attains the pinnacle of accuracy in complex multi-behavior recommendation environments. \\\vspace{-0.12in}

\item Outperforming established disentangled recommendation models such as DGCF and GDCF, our model exhibits exceptional performance across the board. This underscores the essential role of behavior heterogeneity and diversity in accurately mapping the complex mosaic of user behavior patterns for refined recommender systems. At the heart of our model's success is the innovative hierarchical contrastive learning component, which adeptly navigates the intricacies of a heterogeneous hypergraph to deliver this notable enhancement in performance. \\\vspace{-0.12in}


\item Recent efforts like SGL and HCCF use augmentation techniques to address interaction sparsity but mainly focus on homogeneous user-item relationships for computational efficiency. In contrast, we address the sparsity of multi-typed user-item relations by developing an innovative heterogeneous hypergraph contrastive learning framework. This is enhanced by a multi-channel intent representation space that enables adaptive data augmentation. Our approach captures a disentangled representation that incorporates behavior heterogeneity and diversity, offering a deeper understanding of user preferences in recommendations.

\end{itemize}

\begin{figure}
    \centering
    \subfigure[][\scriptsize{Tmall-HR}]{
		\centering
		\includegraphics[width=0.22\columnwidth]{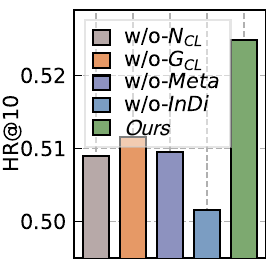}
		\label{fig:ab_ijcai_hr}
	}
	\subfigure[][\scriptsize{Tmall-NDCG}]{
		\centering
		\includegraphics[width=0.22\columnwidth]{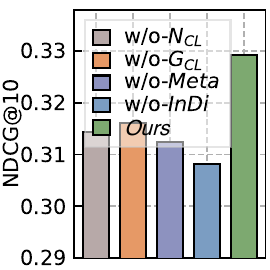}
		\label{fig:ab_ijcai_hr}
	}
    \subfigure[][\scriptsize{IJCAI-HR}]{
		\centering
		\includegraphics[width=0.22\columnwidth]{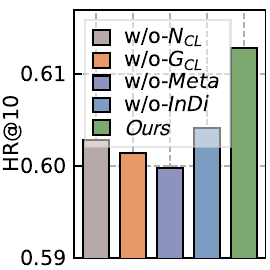}
		\label{fig:ab_ijcai_hr}
	}
	\subfigure[][\scriptsize{IJCAI-NDCG}]{
		\centering
		\includegraphics[width=0.22\columnwidth]{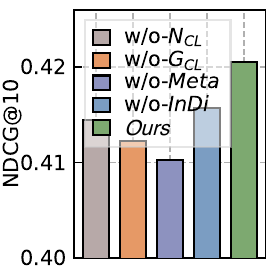}
		\label{fig:ab_ijcai_hr}
	}
	\vspace{-0.15in}
    \caption{Model Ablation Study.}
    \vspace{-0.2in}
    \label{fig:ablation}
\end{figure}

\vspace{-0.05in}
\subsection{Ablation Study of \model\ (RQ2)}
To discern the impact of our \model's core components, we conducted an ablation study with various configurations:

\begin{itemize}[leftmargin=*]
\item \textbf{w/o-$N_{CL}$}: Excluding the node-level adaptive contrastive learning and its data augmentation, this variant omits self-supervised signals that allow for self-discrimination of a user's multi-behavior at the local level among heterogeneous collaborative relations. \\\vspace{-0.12in}

\item \textbf{w/o-$G_{CL}$}: By removing the graph-level multi-behavior contrastive learning, this version ignores the global collaborative context, crucial for understanding the multiplex user-item relationships.

\item \textbf{w/o-$Meta$}: This configuration omits the adaptive contrastive projection mechanism which utilizes a meta network to encode personalized multi-behavior semantics, thereby failing to capture the distinct, individualized behavioral nuances of each user.

\item \textbf{w/o-$InDi$}: This version does not include the intent disentanglement capability offered by the multi-channel parameterized hypergraph structure, which is crucial for creating unique and differentiated representations from diverse behavioral data.

\end{itemize}

The findings presented in Figure~\ref{fig:ablation} clearly demonstrate that the comprehensive \model\ configuration surpasses its modified counterparts, thereby confirming the essential role each component plays in crafting rich representations. The comparison with the w/o-$N_{CL}$ and w/o-$G_{CL}$ variants underscores the significance of hierarchical contrastive learning, which augments user embeddings by incorporating both local and global collaborative contexts from multi-relational interaction data. Furthermore, the superiority of the full \model\ over the w/o-$Meta$ variant showcases its adeptness at adaptively capturing personalized behavior semantics, thus improving model performance via self-supervised augmentation. The enhancements seen with the full \model\ compared to the w/o-$InDi$ variant also validate the model's effectiveness in distilling multi-behavior user representations that acknowledge multiple intents.

\begin{figure}
    \centering
    \subfigure[][HR@10]{
		\centering
		\includegraphics[width=0.42\columnwidth]{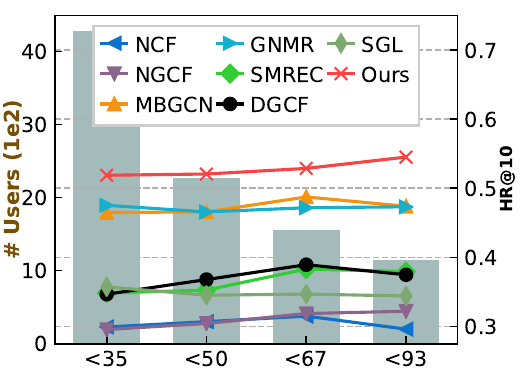}
		\label{fig:sparsity_tmall_hr}
	}
	\subfigure[][NDCG@10]{
		\centering
		\includegraphics[width=0.42\columnwidth]{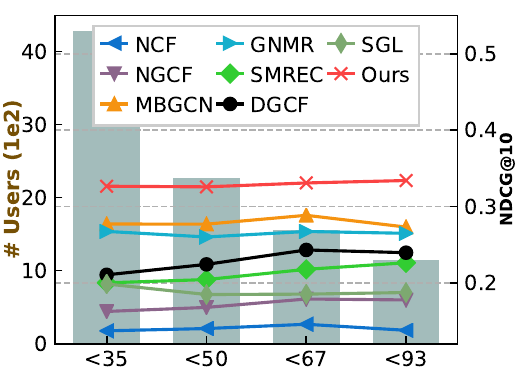}
		\label{fig:sparsity_tmall_ndcg}
	}
    \vspace{-0.15in}
    \caption{Performance \wrt\ user interaction frequency.}
    \label{fig:sparsity}
    \vspace{-0.2in}
\end{figure}

\subsection{Model Resilience Analysis (RQ3)}
We assess the resilience of \model against sparse interaction data in recommender systems. We categorize users into four groups based on interaction frequency (\eg, [35,50), [50,67)) and conduct separate evaluations for each group. Compared to notable baseline models, \model consistently achieves the highest recommendation accuracy across all interaction frequency levels. This performance demonstrates the effectiveness of our model's heterogeneous hypergraph self-augmentation at both node and graph levels, particularly in enhancing recommendations with limited user interaction data. While models like MBGCN and GNMR also leverage diverse user-item interactions to address data scarcity, they face challenges due to sparse supervision of targeted behavior data, leading to suboptimal representations for users and items with long-tail distributions.

\begin{figure*}
    \centering
    \begin{adjustbox}{max width=1.0\linewidth}
    \begin{tikzpicture}
\begin{axis}[
    xlabel={Hidden State Dimensionality d},
    ylabel={HR@10},
    xmin=2,xmax=66,
    ymin=0.40,ymax=0.62,
    legend columns=1,
    legend cell align=right,
    grid=both,
    every axis plot/.append style={ultra thick},
    every tick label/.append style={scale=1.3},
    label style={scale=1.8},
    legend style={
        nodes={scale=1.5, transform shape},
        legend image post style={scale=1.5},
        },
    legend style={at={(1,0)},anchor=south east},
    every axis plot post/.append style={
        every mark/.append style={scale=2}
    }
]
\addplot[color={rgb:red,0;green,157;blue,178}, mark=square, mark options={solid}]
table[x=para, y=tmall_hr] {latFactor-buy.txt};
\addplot[color={rgb:red,245;green,9;blue,11}, mark=triangle, mark options={solid}]
table[x=para, y=ijcai_hr] {latFactor-buy.txt};
\legend{\large Tmall, \large IJCAI};
\end{axis}
\end{tikzpicture}

\begin{tikzpicture}
\begin{axis}[
    xlabel={Hidden State Dimensionality $d$},
    ylabel={NDCG@10},
    xmin=2,xmax=66,
    ymin=0.25,ymax=0.44,
    legend columns=1,
    legend cell align=right,
    grid=both,
    every axis plot/.append style={ultra thick},
    every tick label/.append style={scale=1.3},
    label style={scale=1.8},
    legend style={
        nodes={scale=1.5, transform shape},
        legend image post style={scale=1.5},
        },
    legend style={at={(1,0)},anchor=south east},
    every axis plot post/.append style={
        every mark/.append style={scale=2}
    }
]
\addplot[color={rgb:red,0;green,157;blue,178}, mark=square, mark options={solid}]
table[x=para, y=tmall_ndcg] {latFactor-buy.txt};
\addplot[color={rgb:red,245;green,9;blue,11}, mark=triangle, mark options={solid}]
table[x=para, y=ijcai_ndcg] {latFactor-buy.txt};
\legend{\large Tmall, \large IJCAI};
\end{axis}
\end{tikzpicture}

\begin{tikzpicture}
\begin{axis}[
    xlabel={Number of Intent Hyperedges E},
    ylabel={HR@10},
    xmin=22,xmax=522,
    ymin=0.5,ymax=0.62,
    legend columns=1,
    legend cell align=right,
    grid=both,
    every axis plot/.append style={ultra thick},
    every tick label/.append style={scale=1.3},
    label style={scale=1.8},
    legend style={
        nodes={scale=1.5, transform shape},
        legend image post style={scale=1.5},
        },
    legend style={at={(1,0.3)},anchor=south east},
    every axis plot post/.append style={
        every mark/.append style={scale=2}
    }
]
\addplot[color={rgb:red,0;green,157;blue,178}, mark=square, mark options={solid}]
table[x=para, y=tmall_hr] {hypernum.txt};
\addplot[color={rgb:red,245;green,9;blue,11}, mark=triangle, mark options={solid}]
table[x=para, y=ijcai_hr] {hypernum.txt};
\legend{\large Tmall, \large IJCAI};
\end{axis}
\end{tikzpicture}

\begin{tikzpicture}
\begin{axis}[
    xlabel={Number of Intent Hyperedges $E$},
    ylabel={NDCG@10},
    xmin=22,xmax=522,
    ymin=0.31,ymax=0.43,
    legend columns=1,
    legend cell align=right,
    grid=both,
    every axis plot/.append style={ultra thick},
    every tick label/.append style={scale=1.3},
    label style={scale=1.8},
    legend style={
        nodes={scale=1.5, transform shape},
        legend image post style={scale=1.5},
        },
    legend style={at={(1,0.3)},anchor=south east},
    every axis plot post/.append style={
        every mark/.append style={scale=2}
    }
]
\addplot[color={rgb:red,0;green,157;blue,178}, mark=square, mark options={solid}]
table[x=para, y=tmall_ndcg] {hypernum.txt};
\addplot[color={rgb:red,245;green,9;blue,11}, mark=triangle, mark options={solid}]
table[x=para, y=ijcai_ndcg] {hypernum.txt};
\legend{\large Tmall, \large IJCAI};
\end{axis}
\end{tikzpicture}

\begin{tikzpicture}
\begin{axis}[
    xlabel={Number of GNN Layers L},
    ylabel={HR@10},
    xmin=-0.1,xmax=4.1,
    ymin=0,ymax=0.65,
    legend columns=1,
    legend cell align=right,
    grid=both,
    every axis plot/.append style={ultra thick},
    every tick label/.append style={scale=1.3},
    label style={scale=1.8},
    legend style={
        nodes={scale=1.5, transform shape},
        legend image post style={scale=1.5},
        },
    legend style={at={(1,0)},anchor=south east},
    every axis plot post/.append style={
        every mark/.append style={scale=2}
    }
]
\addplot[color={rgb:red,0;green,157;blue,178}, mark=square, mark options={solid}]
table[x=para, y=tmall_hr] {gnnlayer.txt};
\addplot[color={rgb:red,245;green,9;blue,11}, mark=triangle, mark options={solid}]
table[x=para, y=ijcai_hr] {gnnlayer.txt};
\legend{\large Tmall, \large IJCAI};
\end{axis}
\end{tikzpicture}

\begin{tikzpicture}
\begin{axis}[
    xlabel={Number of GNN Layers $L$},
    ylabel={NDCG@10},
    xmin=-0.1,xmax=4.1,
    ymin=0,ymax=0.46,
    legend columns=1,
    legend cell align=right,
    grid=both,
    every axis plot/.append style={ultra thick},
    every tick label/.append style={scale=1.3},
    label style={scale=1.8},
    legend style={
        nodes={scale=1.5, transform shape},
        legend image post style={scale=1.5},
        },
    legend style={at={(1,0)},anchor=south east},
    every axis plot post/.append style={
        every mark/.append style={scale=2}
    }
]
\addplot[color={rgb:red,0;green,157;blue,178}, mark=square, mark options={solid}]
table[x=para, y=tmall_ndcg] {gnnlayer.txt};
\addplot[color={rgb:red,245;green,9;blue,11}, mark=triangle, mark options={solid}]
table[x=para, y=ijcai_ndcg] {gnnlayer.txt};
\legend{\large Tmall, \large IJCAI};
\end{axis}
\end{tikzpicture}
    \end{adjustbox}
    \vspace{-0.28in}
    \caption{Investigation on the impact of hyperparameters $d, E, L$ for the proposed \model\ framework.}
    \vspace{-0.1in}
    \label{fig:hyperparam}
\end{figure*}
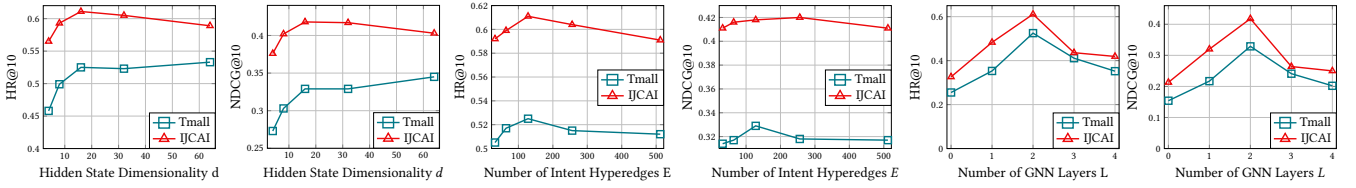

\vspace{-0.05in}
\subsection{Hyperparameter Sensitivity (RQ4)}
Our exploration into the impact of key hyperparameters in \model, such as the dimensionality of hidden states $d$, the count of latent intent hyperedges $E$, and the number of iterations for message passing across both multi-behavior graph and hypergraph structures, has yielded insightful findings. The visualization of these results in Figure~\ref{fig:hyperparam} reveals several patterns shown as follows:

\begin{itemize}[leftmargin=*]

\item (1) Increasing the size of the hidden state dimensionality can unintentionally result in overfitting, especially when working with sparse user and item representations. It requires a delicate balance, as excessive model capacity may capture noise instead of the essential signal for encoding complex user preferences. \\\vspace{-0.12in}

\item (2) When it comes to the number of intent representation channels, our \model\ achieves a sweet spot where performance is optimized. However, beyond this point, there is a decline in performance. This suggests that additional intent hyperedges become unnecessary and may introduce noise that hampers the clarity of the disentangled representation. \\\vspace{-0.12in}

\item (3) Our model achieves optimal performance with a dual-layer message passing structure, effectively utilizing the hypergraph to capture high-order connectivity patterns in heterogeneous interaction data. However, expanding beyond two layers (to three or four) provides no additional benefits and may introduce over-smoothing, which blurs the distinction between diverse user preferences and reduces the model's discriminative power.

\end{itemize}

\begin{figure}[t]
\centering
\includegraphics[width=0.45\textwidth]{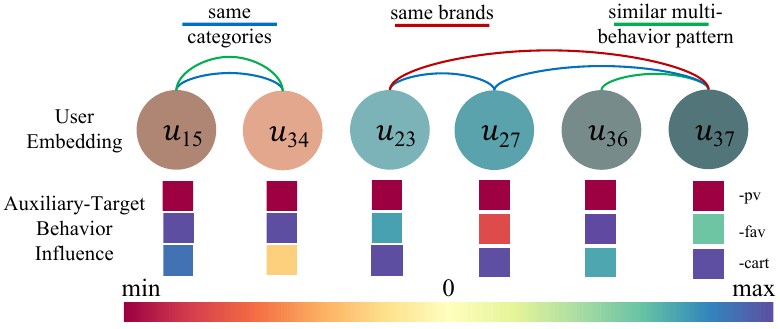}
\vspace{-0.1in}
\caption{Case study of capturing global user dependencies and multi-behavior relationships in latent embedding space.}
\vspace{-0.1in}
\label{fig:case}
\end{figure}

\vspace{-0.05in}
\subsection{Case Study (RQ5)}
To demonstrate our \model's ability to discern global user dependencies and personalized multi-behavioral patterns, we analyze user embeddings, projected and color-coded to indicate similarity, in Figure~\ref{fig:case}. Notably, user pairs ($u_{15}$, $u_{34}$) and ($u_{23}$, $u_{27}$), despite lacking direct connections in the multiplex interaction graph $\mathcal{G}$, exhibit strong interdependence due to similar categories or brands in their product interactions. This suggests our \model's proficiency in capturing collaborative relationships through intent disentanglement and behavior diversity. Additionally, we visualize the relevance weights for auxiliary (\eg, page views) and target behaviors (purchases) within our adaptive meta network, confirming the \model's capability to effectively encode and interpret latent user dependencies and influence across behaviors, underscoring its utility in deciphering complex interaction dynamics.


\vspace{-0.05in}
\subsection{Model Efficiency Study (RQ6)}

To evaluate the efficiency of \model, we compared it against several baselines, assessing computational costs in terms of training and testing times across various datasets. As shown in Table~\ref{tab:time}, \model\ consistently outperforms these baselines, offering shorter training and inference times. Notably, \model\ surpasses KHGT, the top performer in recommendation accuracy, by not only learning more informative user preferences through intent-aware multi-behavior patterns but also by utilizing a more efficient lightweight heterogeneous GNN architecture. Unlike DGCF and HCCF, which focus on a single interaction type, \model\ enhances efficiency by addressing interaction heterogeneity, leveraging its streamlined hypergraph network and graph-level contrastive learning.


\begin{table}[t]
    \centering
    \small
    \caption{Model computational cost with running time (s)}
    \label{tab:time}
    \vspace{-0.1in}
    \begin{tabular}{c|c|c|c|c|c|c}
        \hline
        Data & DGCF & HCCF & KHGT &GNMR & SMRec & \textbf{\model}\\
        \hline
        & \multicolumn{6}{c}{Model Training Time(s)}\\
        \hline
        {Beibei}
        & 11.2 & 15.3 
        & 22.3 & 9.6 & 36.1
        & \textbf{5.8} \\
        \hline
        {Tmall}
        & 37.7 & 29.0 
        & 41.1 & 25.6 & 51.1
        & \textbf{17.2}\\
        \hline
        
        {IJCAI}
        & 43.5 & 35.1 
        & 49.3 & 38.3 & 59.6
        & \textbf{30.4}\\
        \hline
        & \multicolumn{6}{c}{Model Testing Time(s)}\\
        \hline
 
        {Beibei}
        & 9.2 & 7.2 
        & 23.2 & 6.1 & 38.4
        & \textbf{4.5} \\
        \hline
        {Tmall}
        & 30.1 & 32.2 
        & 36 & 24 & 48.7
        & \textbf{21.3}\\
        \hline
        
        {IJCAI}
        & 39.9 & 41.2 
        & 42.3 & 34.5 & 54.5
        & \textbf{37.9}\\
        \hline

        \end{tabular}
        \vspace{-0.1in}
\end{table}

\vspace{-0.05in}
\section{Related Work}
\label{sec:relate}

\noindent \textbf{Multi-Behavior Recommendation}. 
Recent advancements have leveraged diverse user interactions to enhance predictions in recommender systems~\cite{chen2021graph,xuan2023knowledge,yuan2022multi}. Notable models include MBGCN~\cite{mbgcn2020}, a graph convolutional method, KHGT~\cite{xia2020knowledge}, utilizing graph attention, and GNMR~\cite{xia2021multi}, which uses memory-based attention mechanisms for encoding behavior semantics. SMRec~\cite{gu2022self} proposes star-type contrastive regularization for self-supervised signal generation. Despite these developments, none have explored the benefits of learning latent intent factors from heterogeneous relations. This study aims to fill this gap by introducing a parameterized heterogeneous hypergraph to learn disentangled multi-behavior representations.\\\vspace{-0.12in}


\noindent \textbf{Disentangled Representation for Recommendation}. 
Disentangled recommender systems aim to uncover latent factors in complex interaction data~\citep{chen2021curriculum,cao2022disencdr,wang2022learning}. 
DGCF~\citep{ma2019disentangled} facilitates intent-aware disentangled message passing, while MacridVAE~\citep{ma2019learning} employs a variational auto-encoder to disentangle user representations. GDCF~\citep{zhang2022geometric} uses non-Euclidean frameworks to disentangle user-item interactions, inspired by hybrid geometries representation learning. DcRec~\citep{wu2022disentangled} introduces disentangled contrastive learning in social recommendations. Despite these innovations, current methods do not fully address the heterogeneity of user behaviors, a significant challenge for disentangled collaborative filtering.\\\vspace{-0.12in}


\noindent \textbf{Contrastive Recommendation Models}.
Contrastive Learning (CL) has become a key method in recommendation~\cite{wei2022contrastive}, improving both graph-based and sequential recommendation. SGL~\citep{wu2021self} and RecDCL~\citep{zhang2024recdcl} refine user and item representations through self-discrimination tasks. In sequential recommendations, CL4SRec~\citep{qiu2022contrastive} and DuoRec~\citep{xie2022contrastive} enhance sequence representations using techniques like masking, cropping, and reordering. CL is also applied in various areas, including micro-video recommendation~\citep{wei2021contrastive}, social-aware recommendation~\citep{li2024recdiff}, and group recommendation~\citep{li2023self}.


\vspace{-0.05in}
\section{Conclusion}
\label{sec:conclusion}
This paper introduces \model, addressing heterogeneous disentangled collaborative filtering by learning factorized representations that separate latent intents in user interactions. We enhance the embeddings' expressiveness and robustness with a hierarchical contrastive learning method, using adaptive augmentation through parameterized heterogeneous hypergraphs. Our experiments on public datasets show \model's superior performance. Future work will explore pre-training with heterogeneous hypergraphs to further improve representations for diverse recommender systems.

\clearpage
\bibliographystyle{abbrv}
\balance
\bibliography{refs}
\clearpage
\section*{Ethical Considerations}

In this section, we explore the ethical implications of the \model\ model, which aims to disentangle heterogeneous interaction patterns between users and items to enhance recommendation accuracy. We will examine and discuss potential ethical challenges that may arise during the implementation and operation of the \model. \\\vspace{-0.12in}

\noindent\textbf{Privacy and Data Security}.
The \model\ framework, like other collaborative filtering systems, relies heavily on extensive user data to personalize content recommendations, raising significant privacy concerns for all stakeholders involved. To protect user information from unauthorized access and potential breaches, it is essential to implement comprehensive data security measures, including robust encryption practices, secure data storage solutions, and regular security audits. Additionally, transparency regarding the usage and handling of user data is crucial, ensuring that users are fully informed and have consented to the utilization of their information. Techniques such as data anonymization and differential privacy should be employed to minimize the risk of re-identifying individuals from collected data, thereby safeguarding the confidentiality and security of user activities within the system. \\\vspace{-0.12in}


\noindent\textbf{Bias and Fairness}.
Inherent biases in the data used to train collaborative filtering algorithms, such as those in the \model\ framework, can perpetuate and amplify these biases, potentially leading to unfair outcomes for specific user groups. Addressing these biases is crucial and necessitates the implementation of fairness-enhancing techniques, including the development of models sensitive to demographic parity and equality of opportunity. Regular audits of recommendation outputs are essential to identify, monitor, and rectify any emerging biases. Additionally, adjusting algorithms to promote the visibility of underrepresented or niche content helps mitigate the dominance of popular items in recommendations. These measures ensure a balanced representation of diverse content, fostering a more inclusive and equitable platform for all users.


\end{document}